\documentclass{svjour3}                     
\smartqed  
\usepackage{graphicx}
\begin{document}

\title{First exclusive measurements of the $K^-pp$ state\\
populated in the $pp \rightarrow K^+ \Lambda p$ reaction at 2.85 GeV
}


\author{
T.~Yamazaki \and
P.~Kienle \and
K.~Suzuki \and
M.~Maggiora on behalf of DISTO Collaboration 
(M.~Alexeev \and
F.~Balestra \and
Y.~Bedfer
\and
R.~Bertini \and
L.~C.~Bland
\and
A.~Brenschede
\and
F.~Brochard \and
M.~P.~Bussa \and
M.~Chiosso \and
Seonho~Choi
\and
M.~L.~Colantoni \and
R.~Dressler
\and
M.~Dzemidzic
\and
J.-Cl.~Faivre \and
L.~Ferrero \and
J.~Foryciarz
\and
I.~Fr\"ohlich
\and
V.~Frolov
\and
R.~Garfagnini \and
A.~Grasso \and
S.~Heinz
\and
W.~W.~Jacobs \and
W.~K\"uhn \and
A.~Maggiora \and
D.~Panzieri \and
H.-W.~Pfaff \and
G.~Pontecorvo \and
A.~Popov \and
J.~Ritman
\and
P.~Salabura \and
V.~Tchalyshev \and
F.~Tosello \and
S.~E.~Vigdor
\and
G.~Zosi)\\
}

\institute{
T.~Yamazaki \at Department of Physics, University of Tokyo, Tokyo, Japan and RIKEN Nishina Center, Saitama, Japan\\
\email{yamazaki@nucl.phys.s.u-tokyo.ac.jp}
\and
P.~Kienle and K.~Suzuki \at Stefan Meyer Institute for Subatomic Physics, Austrian Academy of Sciences, Vienna, Austria
\and
P.~Kienle \at Excellence Cluster Universe, Technische Universit\"at M\"unchen, Garching, Germany\and
M.~Alexeev, F.~Balestra, R.~Bertini, M.~P.~Bussa, M.~Chiosso, M.~L.~Colantoni, A.~Ferrero, L.~Ferrero, R.~Garfagnini, A.~Grasso, S.~Heinz, A.~Maggiora, M.~Maggiora, G.~Pontecorvo, F.~Tosello and G.~Zosi
\at Dipartimento di Fisica Generale "A. Avogadro" and INFN, Torino, Italy
\and
Y.~Bedfer, R.~Bertini, F.~Brochard, J.-Cl.~Faivre and S.~Heinz \at Laboratoire National Saturne, CEA Saclay, France
\and
L.~C.~Bland, Seonho~Choi, M.~Dzemidzic, W.~W.~Jacobs and S.~E.~Vigdor \at Indiana University Cyclotron Facility, Bloomington, Indiana, U.S.A.
\and
A.~Brenschede, I.~Fr\"ohlich, W.~K\"uhn, H.-W.~Pfaff and J.~Ritman \at II. Physikalisches Institut, Univ. Gie\ss{}en, Germany
\and
R.~Dressler \at  Forschungszentrum Rossendorf, Germany
\and
J.~Foryciarz and P.~Salabura \at M. Smoluchowski Institute of Physics, Jagellonian University, Krak\'ow, Poland 
\and
J.~Foryciarz \at H.~Niewodniczanski Institute of Nuclear Physics, Krak\'ow, Poland
\and
V.~Frolov, G.~Pontecorvo and A.~Popov and V.~Tchalyshev \at JINR, Dubna, Russia
\and
D.~Panzieri \at Universit\`a del Piemonte Orientale and INFN, Torino, Italy
}


\date{Prepared for EXA08: October 27, 2008 / Accepted: date}

\maketitle

\begin{abstract}
We have analyzed data of the DISTO experiment on the exclusive $pp \rightarrow K^+ \Lambda p$ process at $T_p = 2.85$ GeV to search for a $K^-pp ~(\equiv X)$ nuclear bound state to be formed in the $pp \rightarrow K^+ + X$ reaction. The deviation spectra of the $K^+$ missing-mass $\Delta M (K^+)$ and $\Lambda p$ invariant-mass $M(\Lambda p)$ with selection of large-angle proton emission revealed a structure with $M_X = 2265 \pm 2$ MeV/$c^2$ and $\Gamma_X = 118 \pm 8$ MeV.

\end{abstract}


\section{Introduction}
The simplest kaonic nuclear bound system, $K^-pp$, which was first predicted to be a quasi-stable state with $M = 2322$ MeV/$c^2$, $B_K = 48$ MeV and $\Gamma_{\Sigma \pi p} = 61$ MeV \cite{Akaishi:02,Yamazaki:02}, has been studied in recent  years. A detailed theoretical analysis, based on the ansatz that the $\Lambda (1405)$ resonance (hereafter called $\Lambda^*$) is an $I=0$ $\bar{K} N$ quasi-bound state embedded in the $\Sigma \pi$ continuum, has shown that $K^-pp$ has a molecule-like structure in which the $K^-$ migrates between the two protons, causing a {\it super-strong nuclear force} \cite{Yamazaki:07a,Yamazaki:07b}. The strongly bound nature of $K^-pp$ is also supported by recent Faddeev calculations \cite{Shevchenko,Ikeda}. On the other hand, various theories based on chiral dynamics lead to a ``weak" $\bar{K} N$ interaction and a shallow bound state \cite{Oset,Weise:07}. Since the issue is pertinent to the problem of kaon condensation, it is of vital importance to distinguish between the ``strong binding" and the ``weak binding" regimes by studying $K^-pp$ experimentally, but only little information is known from a FINUDA experiment of $K^-$ absorption at rest in light nuclei \cite{Agnello:05}. 
In the mean time, it was predicted that the strongly bound $K^-pp$ system with a short $p-p$ distance can be formed in a $p+p \rightarrow K^+ + K^-pp$ reaction with an enormously large sticking probability between $\Lambda^*$ and $p$, which is brought about due to the short range and large momentum transfer of the $pp$ reaction \cite{Yamazaki:07b}. Here, we report that existing experimental data of $p + p \rightarrow K^+ + \Lambda + p$ taken by the DISTO spectrometer show an evidence for this exotic formation.

\section{Dalitz plot of observed $p \Lambda K^+$ events at 2.85 GeV.}

The DISTO experiment, originally aimed at comprehensive studies of  polarization transfer in strangeness exchange reactions in $pp$ collisions, was carried out by using a proton beam (up to 2.85 GeV) from the SATURNE accelerator at Saclay \cite{DISTO-NIM,DISTO}. Here, we have analyzed the experimental data set of the exclusive reaction channel, 
\begin{equation}
 p + p \rightarrow p + \Lambda + K^+, \label{eq:pp2LpK}
\end{equation}
at the incident energy of 2.85 GeV. 
The events in this channel were first selected by the $\Lambda$ particle identification from the $p \pi^-$ invariant-mass spectrum, and then from a missing-mass spectrum of $p K^+$, $\Delta M (p K^+)$, which shows peaks of $\Lambda$, $\Sigma^0$ and $\Sigma^0(1385)/\Lambda(1405)$. Here, we report on the $p \Lambda K^+$ data of about 140,000 events.

\begin{figure}[t]
  \begin{center}
    \parbox[c]{0.47\textwidth}{
      \centering
      \includegraphics[width=0.47\textwidth]{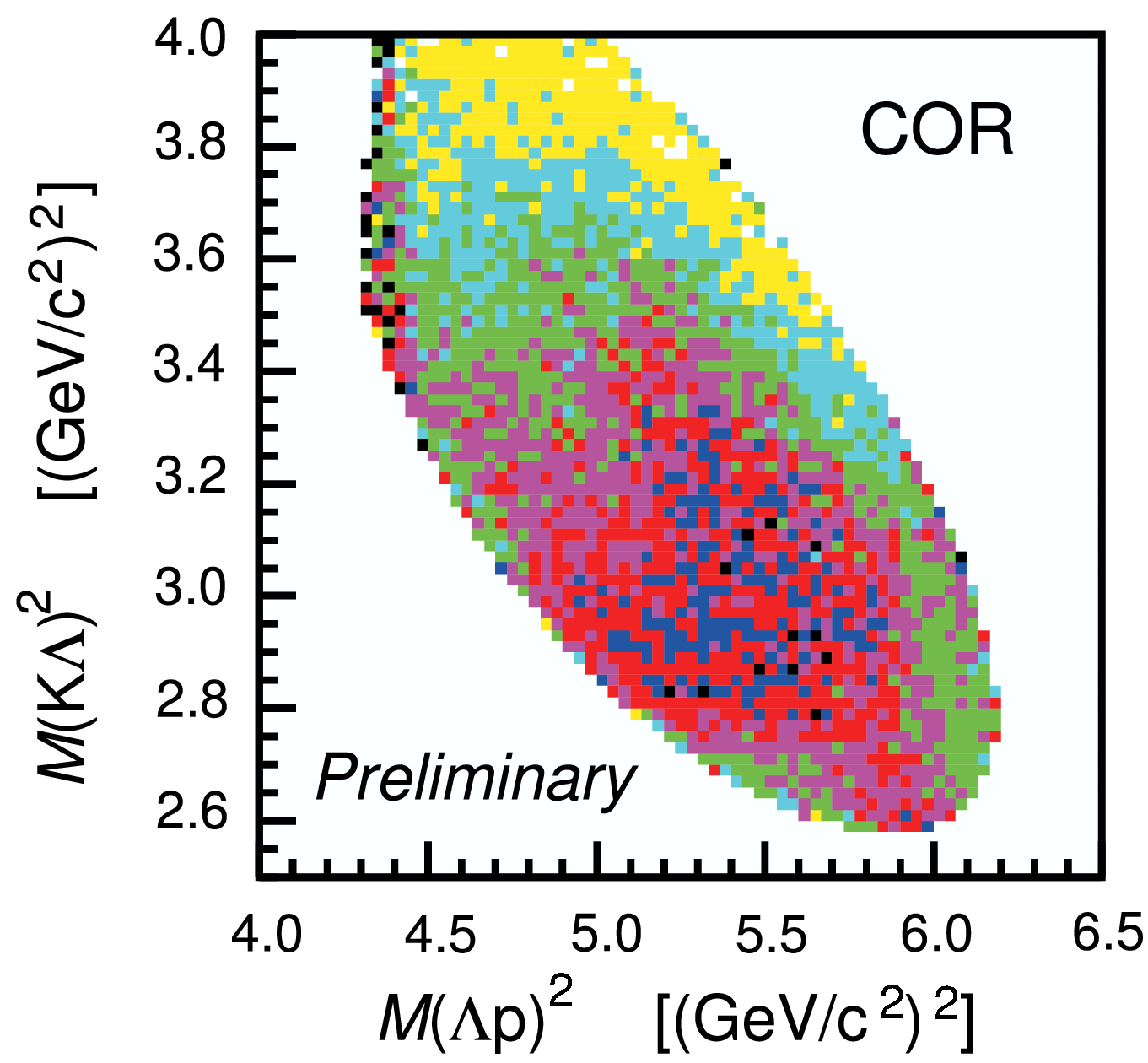}
    }
    \parbox[c]{0.51\textwidth}{
      \caption{
        An acceptance corrected Dalitz plot, $M(\Lambda p)^2$ vs $M(K\Lambda)^2$, of $pp \rightarrow p \Lambda K^+$ at 2.85 GeV.
      }
      \label{fig:DISTO-Dalitz}
    }
  \end{center}
\end{figure}

The data at 2.85 GeV cover the kinematical region in which we can look for a candidate of the $K^-pp$  bound state ($\equiv X$):
\begin{equation}\label{eq:pp2KX}
p + p \rightarrow K^+  + X,~~~
         X \rightarrow p + \Lambda.
\end{equation}
 The exotic process leading to a two-body final state ($K^+ X$) is a part of the whole $p \Lambda K^+$ process, which consists of the ordinary  process (hereafter called {\it background process}) and the exotic process. We show in Fig.~\ref{fig:DISTO-Dalitz} an acceptance-corrected Dalitz plot of  all the $p \Lambda K^+$ events in the plane of $x = M(\Lambda p)^2$~ vs~ $y = M(K\Lambda)^2$. The Dalitz distribution calculated by taking into account the reaction dynamics \cite{Akaishi;08b} is different from the "uniform" phase-space distribution, but {\it is continuous without any local bump structure}. On the other hand, the observed distribution, Fig.~\ref{fig:DISTO-Dalitz}, reveals some structure that cannot be explained by the ordinary process.
 Now, we proceed to studies of the angular  correlations of the three particles, $p$, $\Lambda$ and $K^+$, which are hidden in the kinematical variables in the Dalitz presentation. 

\begin{figure}[htbp]
\centering
\includegraphics[width=12cm]{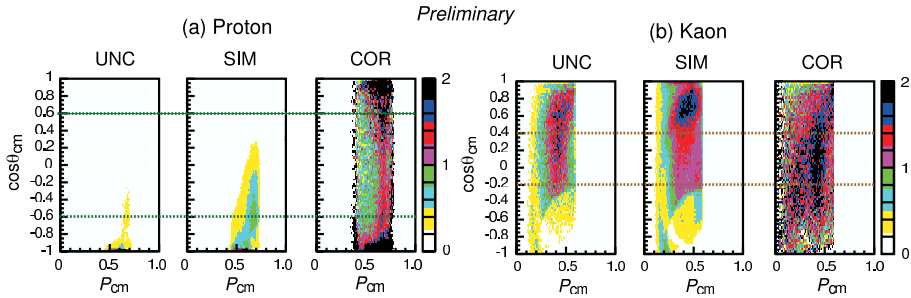}
\vspace{0cm}
\caption{\label{fig:p-K-distribution} 
Momentum distributions, $P$, versus cos$ \, \theta$ in c.m. of (a) $p$ and (b) $K^+$. Each block consists of  $UNC$ (uncorrected), $SIM$ (simulated), $COR$ (corrected) data. The horizontal dotted lines zone $-0.6 < {\rm cos}\, \theta_{\rm cm} (p) < 0.6$ and $-0.2 < {\rm cos} \, \theta_{\rm cm}(K) < 0.4$, to be used for proton-angle and $K^+$-angle cuts. }
\end{figure}

\section{Angular correlations of $p \Lambda  K^+$}

Figure \ref{fig:p-K-distribution} shows the uncorrected ($UNC$), simulated ($SIM$) and corrected ($COR$)  distributions of the momentum $P$ vs cos$\,\theta$ in c.m. of $p$ and $K^+$. Since the $SIM$ data are calculated by taking into account the detector acceptance for events of uniform phase-space distribution, 
the corrected data here represent, {\it not} the spectral intensities, but ``{\it deviation spectra}" $DEV$, which indicate how much the data deviate from the expectation based on the uniformly distributed simulation events, namely, $COR = UNC/SIM \rightarrow DEV$. The uncorrected data as well as the simulation data show that the $p$ distribution is extremely backward, corresponding to the extremely forward $\Lambda$ distribution, which arises from the large acceptance of the DISTO detector for $\Lambda \rightarrow p + \pi^-$ in the forward direction. Nevertheless, the corrected distributions of both $p$ and $\Lambda$ are remarkably symmetric, as expected from the symmetric $pp$ collision in c.m., which justifies the present acceptance correction. Since the proton group at extremely backward angles (cos$ \, \theta_{\rm cm} (p) < -0.2$) is related to low-momentum transfer ($q <0.3$ GeV/$c$) production of $p \Lambda K^+$, it is likely to correspond to the background process. On the other hand, the proton group of large-angle emission corresponds to the exotic process involving the decay of $X$ with a transverse momentum of around 0.5 GeV/c. So, it is extremely interesting to distinguish them by applying proton-angle cuts.

\begin{figure}[htbp]
\centering
\includegraphics[width=12cm]{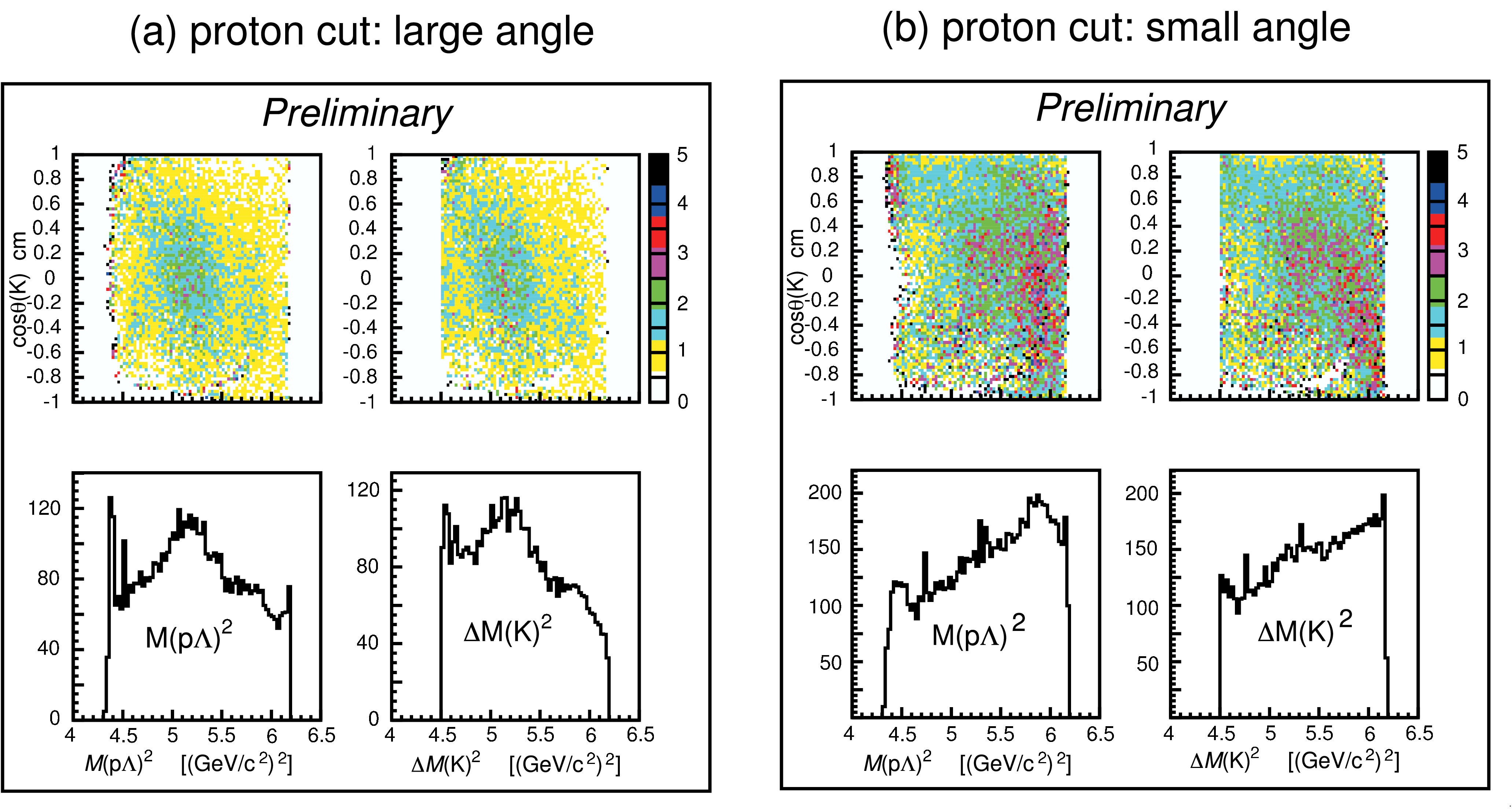}
\vspace{0cm}
\caption{\label{fig:Dalitz+K-pcut} 
(a) Events with large-angle proton cut ($-0.6 < {\rm cos}\, \theta _{\rm cm} (p) < 0.6$) and (b) events with small-angle proton cut ($|{\rm cos}\, \theta _{\rm cm} (p)| > 0.6$). Each frame consists of (upper) deviation spectra of $M (p \Lambda)^2$ and $\Delta M (K)^2$ vs cos$\, \theta _{\rm cm} (K)$ and (lower) their projections. 
}
\end{figure}

Figure.~\ref{fig:p-K-distribution} (b) already shows that the $P_{\rm cm} (K)$ has a monoenergetic component around 0.4 GeV/c even before applying the proton-angle cut, which must be a signature for two-body exotic final states, $K^+ X$. It is to be noted that this component is present in the uncorrected data, and the component appearing in the corrected data {\it cannot be a fake} that might originate from the correction, since $SIM$ is smooth in this region. Figure~\ref{fig:Dalitz+K-pcut} (upper) shows deviation spectra of $M (p \Lambda)^2$ and $\Delta M (K)^2$ vs cos$\, \theta _{\rm cm} (K)$ and Fig.~\ref{fig:Dalitz+K-pcut} (lower) their projections with (a) large-angle protons and (b) small-angle protons.
  
Clearly, the vertical band in the $P_{\rm cm}(K)$ vs cos$\, \theta_{\rm cm} (K)$ plot (Fig.~\ref{fig:p-K-distribution} (b)) is enhanced with a large-angle proton cut, as shown in Fig.~\ref{fig:Dalitz+K-pcut}, where the $P_{\rm cm}(K)$ spectra are converted into missing-mass deviation spectra, $\Delta M (K)^2$. The deviation spectra of both $M(p\Lambda)^2$ and $\Delta M (K)^2$ in Fig.~\ref{fig:Dalitz+K-pcut} (a) show an enhanced peak at around $x \sim 5.15$, corresponding to $M_X \approx 2.27$ GeV/$c^2$.

\begin{figure}[htbp]
\centering
\includegraphics[width=12cm]{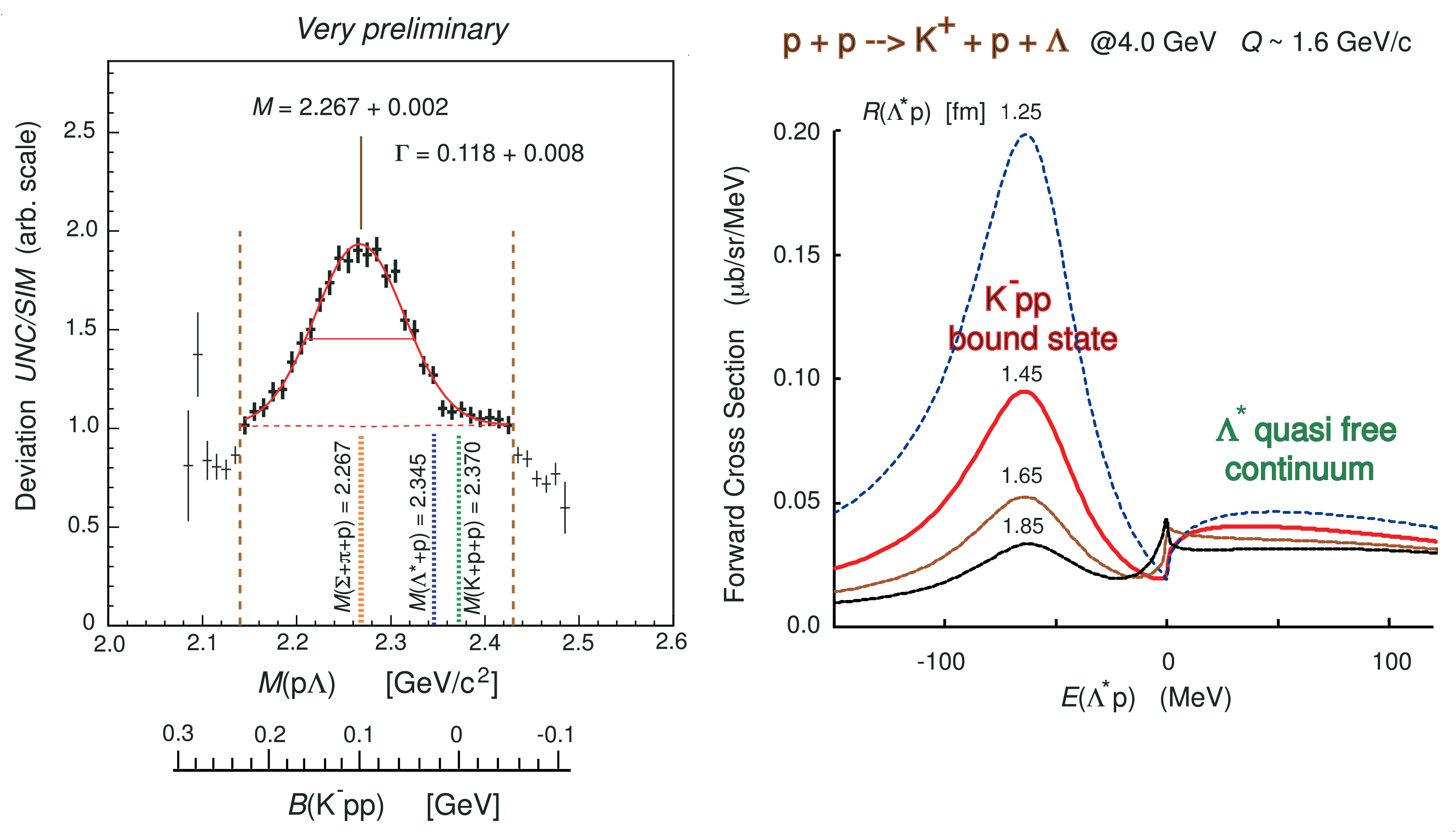}
\vspace{0cm}
\caption{\label{fig:MM-K} 
(Left) A final deviation spectrum of $M(p \Lambda)$ after selection of large-angle proton emission: $-0.6 < {\rm cos} \, \theta_{\rm cm} (p) < 0.6$ and large-angle $K^+$ emission, $-0.2 < {\rm cos} \, \theta_{\rm cm} (K^+) < 0.4$. The peak structure with a background (thin line) gives $M_X$ and $\Gamma_X$. The three important masses of $K^- + p + p = 2.370$, $\Lambda^* + p = 2.345$, and $\Sigma^+ + \pi^- + p = 2.267$ GeV/$c^2$ are indicated by vertical dotted lines. The safe zone of acceptance is also shown by two vertical brown broken lines. (Right) Calculated forward cross sections of the $p(p,K^+)K^-pp$ reaction at 4 GeV for different rms distances $R(\Lambda^* p)$. The cases of $R(\Lambda^* p)$ = 1.65 and 1.45 fm correspond to the $\bar{K}N$ interactions of the original Akaishi-Yamazaki version ($B_K = 48$ MeV) and the version with 17\% enhancement  ($B_K = 86$ MeV) \cite{Yamazaki:07b}, respectively.
}
\end{figure}

We have demonstrated that the proton-angle cut is very effective in discriminating the $X$ formation process (monoenergetic emission of $K^+$) from the background process. On the other hand, the small-proton-angle events correponding to small-$q$ transfer can be attributed to the background process. In fact, the deviation spectra, $M(p \Lambda)$ and $\Delta M (K^+)$, of small proton-angle events, as shown in Fig.~\ref{fig:Dalitz+K-pcut} (b), are remarkably flat, in great contrast to those of large proton angle events (a). The flat shape of (b) is well accounted for by a $pp \rightarrow p \Lambda K^+$ mechanism with a collision length of $\hbar/m_B c$ with $m_B \sim 0.4$ GeV/$c^2$ 
\cite{Akaishi;08b}. In great contrast, the deviation spectra of large-angle proton emission (a) are far from a flat shape, revealing a large bump.

We also find another important fact from the $P_{\rm cm}(K)$ vs cos$ \, \theta_{\rm cm} (K)$ distribution; that is, the monoenergetic $X$-events are enhanced in the large-angle $K^+$ emission. We also notice from Fig.~\ref{fig:p-K-distribution} (b) that the zone of ${\rm cos} \, \theta_{\rm cm} (K) < -0.2$ in both $UNC$ and $SIM$ is very much depleted. So, we applied cuts: $-0.2 < {\rm cos} \, \theta_{\rm cm} (K) < 0.4$, to obtain final deviation spectra, Fig.~\ref{fig:MM-K}.

\section{Concluding Remarks}

 We have observed a large broad peak, which is associated with protons of large transverse momenta. Such a peak is not seen at all in the deviation spectra of small proton-angle events. Thus, we conclude that this structure is most likely due to a $K^-pp$ bound state formed in the $pp \rightarrow K^+ + X$ reaction followed by the decay $X \rightarrow p + \Lambda$. We made simple fitting of the deviation spectra and obtained 
\begin{eqnarray}
&& M_X = (2.265 \pm 0.002)~{\rm GeV}/c^2,\\
&& \Gamma_X  = (0.118 \pm 0.008)~{\rm GeV}/c^2,
\end{eqnarray}
when a Gaussian shape on a linear background was assumed. The best-fit $\chi ^2 /ndf$ value is 34.2/24 = 1.4, whereas the $\chi ^2 /ndf$ value without Gaussian peak becomes 946.6 /27 = 35.1.  The peak height amounts to about 26 $\sigma$.
The mass of $X$ corresponds to $B_K = (105 \pm 2)$ MeV for $X = K^-pp$. This is close to the mass $M(\Lambda p) \sim 2.256$ GeV/$c^2$ of the claimed $K^-pp$ candidate reported in the stopped-$K^-$ experiment by FINUDA \cite{Agnello:05}. The mass spectrum is compared with some important particle-emission thresholds: $M(K^- + p + p) = 2.370$ GeV/$c^2$, $M(\Lambda (1405) + p) = 2.345$ GeV/$c^2$, and $M(\Sigma^+ + \pi^- + p) = 2.267$ GeV/$c^2$. 

It is to be stressed that the large population of a $K^- pp$ state observed in the $pp$ reaction provides a direct evidence for the unusually high-density regime of the kaonic nuclear system, as demonstrated by the comparison between the observed spectrum and the predicted ones in Fig.~\ref{fig:MM-K}. The observed binding energy is larger than the original prediction, thus suggesting additional effects to be pursued \cite{Yamazaki:07b,Wycech}. On the other hands, some theoretical claims for shallow $\bar{K}$ binding \cite{Oset,Weise:07} do not seem to be reconciled with the present observation. The observed structure calls for theoretical studies of the decay shape \cite{Akaishi:08} and branch \cite{Ivanov:07} of $K^- pp$.\\

This research was partly supported by the DFG cluster of excellence ``Origin and Structure of the Universe" and by Grant-in-Aid for Scientific Research of Monbu-Kagakushou of Japan. One of us (T.Y.) acknowledges the support by an Award of the Alexander von Humboldt Foundation, Germany.

\end{document}